\documentclass[fleqn,usenatbib]{rasti}

\usepackage{newtxtext,newtxmath}

\usepackage[T1]{fontenc}

\DeclareRobustCommand{\VAN}[3]{#2}
\let\VANthebibliography\thebibliography
\def\thebibliography{\DeclareRobustCommand{\VAN}[3]{##3}\VANthebibliography}

\usepackage{graphicx}	
\usepackage{aas_macros}
\usepackage{siunitx}
\usepackage{booktabs}

\def\noopsort#1{}
\DeclareSIUnit{\sqrthz}{\ensuremath{\sqrt{\text{\hertz}}}}
\DeclareSIUnit{\vrms}{\volt_{RMS}}

\title[BiSON detector bandwidth and polarisation]{Detector bandwidth
  and polarisation switching rates: spectrophotometric observations of
  the Sun by the Birmingham Solar Oscillations Network (BiSON)}

\author[S. J. Hale et al.]{
S. J. Hale,$^{1}$\thanks{E-mail: s.j.hale@bham.ac.uk}
W. J. Chaplin,$^{1}$
G. R. Davies,$^{1}$
Y. P. Elsworth,$^{1}$
R. Howe,$^{1}$
\\
$^{1}$School of Physics and Astronomy, University of Birmingham, Edgbaston, Birmingham B15 2TT, United Kingdom\\
}

\date{Accepted XXX. Received YYY; in original form ZZZ}

\pubyear{2022}

\begin{document}
\label{firstpage}
\pagerange{\pageref{firstpage}--\pageref{lastpage}}
\maketitle

%
%
%
%


\begin{abstract}
The Birmingham Solar Oscillations Network (BiSON) observes acoustic
oscillations of the Sun.  The dominant noise source is caused by
fluctuations of Earth's atmosphere, and BiSON seeks to mitigate this
effect by combining multiple rapid observations in alternating
polarisation states.  Current instrumentation uses bespoke
Pockels-effect cells to select the polarisation state.  Here, we
investigate an alternative off-the-shelf solution, a liquid crystal
retarder, and discuss the potential impact of differences in
performance.  We show through electrical simulation of the
photodiode-based detectors, and assessment of both types of
polarisation device, that although the switching rate is slower the
off-the-shelf LCD retarder is a viable replacement for a bespoke
Pockels-effect cell.  The simplifications arising from the use of
off-the-shelf components allows easier and quicker instrumentation
deployment.
\end{abstract}




%
%
%
%


\section{Introduction}
\label{s:introduction}

The Birmingham Solar Oscillations Network (BiSON) is a ground-based
network of solar observatories, consisting of six sites measuring
acoustic oscillations of the Sun~\citep{2016SoPh..291....1H,halephd}.
The instrumentation makes high-precision measurements of the Doppler
velocity of the solar surface by making use of spectrophotometry --
photometry of a spectral line.  BiSON observes the D1~transition of
potassium, which has a central wavelength
of~\SI{769.898}{\nano\meter}.

Photometry from the ground is challenging due to fluctuation of
Earth's atmosphere, known as atmospheric scintillation, which
typically becomes the dominant noise source. In order to limit the
effect of this source of noise it is necessary to restrict detector
exposure times to
generally~\SI{<10}{\milli\second}~\citep{2015MNRAS.452.1707O}.
Measurement of scintillation at the BiSON sites has shown that
the~\SI{-3}{\decibel} point typically occurs at around~\SI{5}{\hertz},
and~\SI{-10}{\decibel} occurs
above~\SI{20}{\hertz}~\citep{2020PASP..132c4501H}.

The BiSON spectrophotometers mitigate this problem by rapidly
switching the measurement wavelength between the red and the blue
wings of the solar absorption line, and then normalising by the sum of
the intensities at the two working points to form an intensity ratio,
$R$,
\begin{equation}
  \label{eq:ratio}
  R = \frac{I_{\mathrm{b}} - I_{\mathrm{r}}}{I_{\mathrm{b}} + I_{\mathrm{r}}} ~,
\end{equation}
where $I_{\mathrm{b}}$ and $I_{\mathrm{r}}$ are the intensities
measured at the blue and red wings of the solar absorption line,
respectively.  It is this ratio measurement that is then integrated
over an acquisition period of~\SI{4}{\second}.  The typical
instrumental velocity sensitivity is~\SI{3000}{\meter\per\second} per
unit ratio~\citep{1995A&AS..113..379E}.  The level of reduction of
scintillation noise is dependent on the switching rate, where higher
switching rates produce better noise suppression but are more
difficult to achieve.

Wavelength selection is controlled through polarisation switching.
Both polarisation states are multiplexed through a single detector in
order to avoid differences in noise, gain, or sensitivity.  The
detector has both a high gain requirement, and also a relatively high
bandwidth requirement (in comparison to the few~\si{\milli\hertz}
solar oscillations) due to the transient signal changes. The original
detectors used in BiSON instrumentation in the early 1990s were
designed and adjusted empirically to achieve optimisation.  In
Section~\ref{s:spice} we take a formal approach to performance
analysis through circuit simulation using SPICE, the industry standard
electrical simulation tool, to determine detector noise levels,
frequency response, and slew rate.  In Section~\ref{s:polarisation} we
look at methods of electronically controlling light polarisation
state.  Existing BiSON instrumentation uses bespoke field-widened
Pockels-effect cells.  Many solar Zeeman spectropolarimetry projects
have moved to using off-the-shelf Nematic Liquid Crystal (NLC)
variable retarders in order to improve both switching rate and
mechanical reliability (see,~e.g.,~\citealt{2006spse.conf...37Z,
  2010MmSAI..81..763J, 2010A&A...520A.115B, 2010PASP..122..420H,
  2011SoPh..268...57M, 2017RAA....17....8G, 2019JATIS...5c4002P,
  2020SoPh..295..109R}).  However, NLCs are typically much slower to
transition between states than Pockels-effect cells.  A Ferro-electric
Liquid Crystal (FLC) polarisation modulator~\citep{10.1117/1.602183,
  10.1117/12.857045, 10.1117/12.2305297, 10.1117/1.OE.58.8.082417} can achieve
switching rates comparable to a Pockels-effect cell, but has the
disadvantage of being considerably more expensive than NLCs.  Here, we
look at only the more affordable NLCs as potential replacements for
our Pockels-effect cells.  Finally, in Section~\ref{s:discussion}, we
discuss the impact on BiSON spectrophotometer design and
configuration~\citep{10.1093/rasti/rzac007}, and the move from
Pockels-effect cell to LC-based polarisation control.


%
%
%
%


\section{Detector Simulation}
\label{s:spice}

\begin{figure}
  \includegraphics[width=\columnwidth]{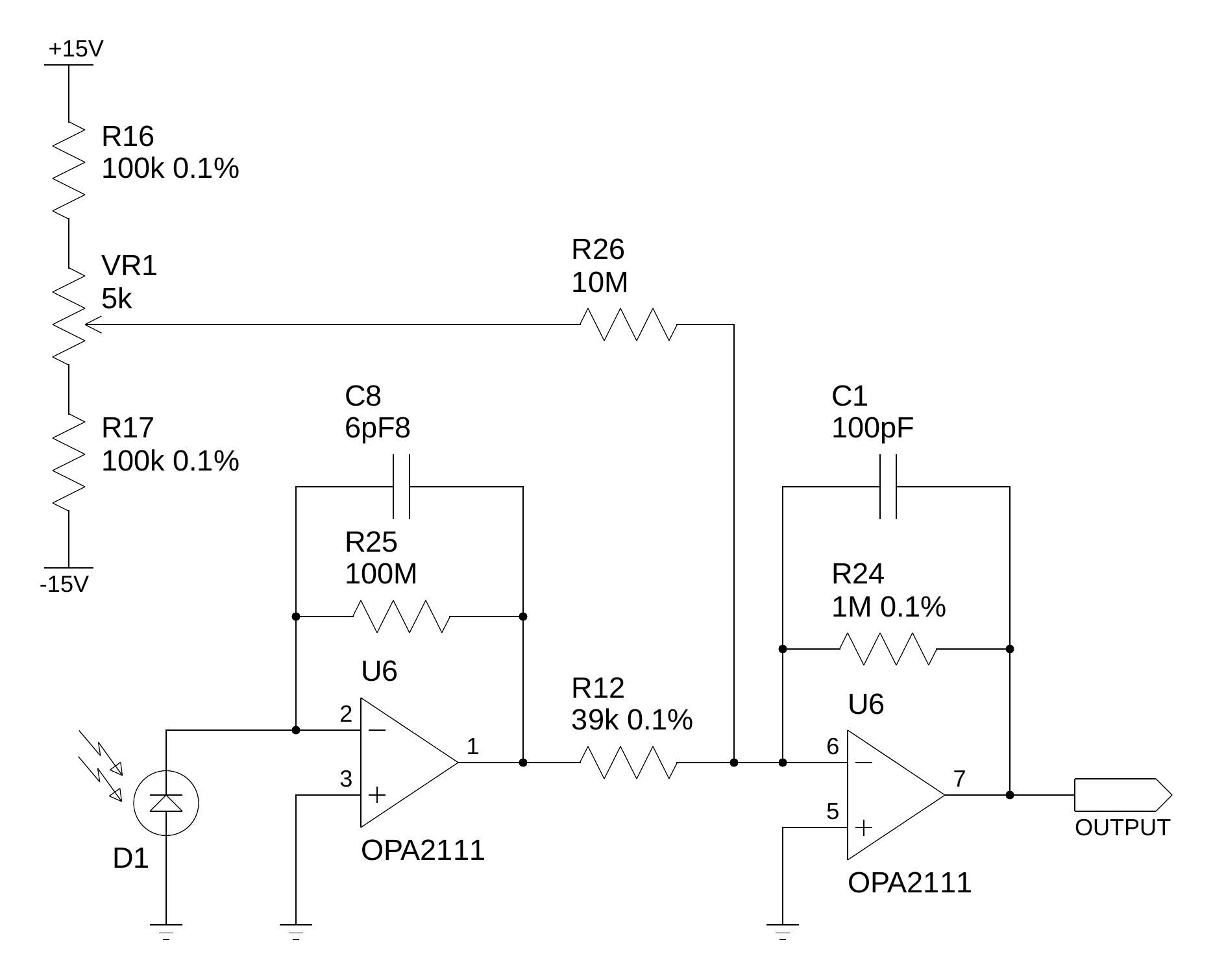}
  \caption{Transimpedance amplifier stage.}
  \label{fig:transimpedance}
\end{figure}

\begin{figure}
  \includegraphics[width=\columnwidth]{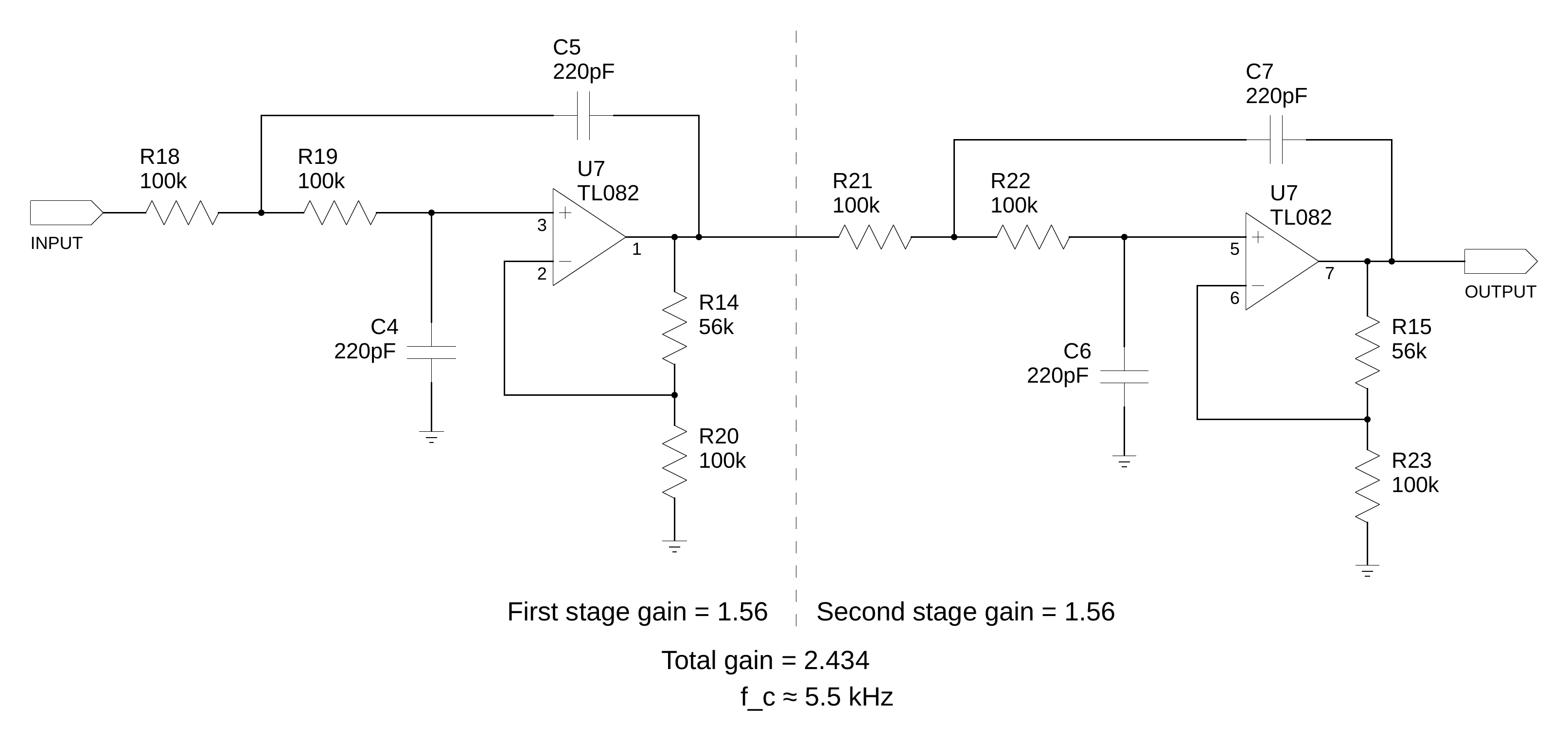}
  \caption{Low-pass Bessel filter stage.}
  \label{fig:bessel}
\end{figure}

The detector schematic is shown in Figure~\ref{fig:transimpedance}.
The circuit uses a transimpedance amplifier for initial
photocurrent-to-voltage conversion, followed by a second stage voltage
amplifier for additional gain.  The scattering detectors receive
between~\SI{0.35}{\nano\watt} and~\SI{1.4}{\nano\watt} of light
depending on atmospheric conditions and the line-of-sight velocity
offset of the spectral line (\citealt{halephd};
\citealt{10.1093/rasti/rzac007}).  The transimpedance amplifier requires
a~\SI{100}{\mega\ohm} gain resistor to convert a few
hundred~\si{\pico\ampere} of photocurrent to an output of a
few~\si{\milli\volt}.  The second stage boosts this to a few
hundred~\si{\milli\volt}.  The signal is then passed through a 4-pole
low-pass Bessel filter, shown in Figure~\ref{fig:bessel}, which
removes high-frequency noise and also provides a little more gain to
boost the output dynamic range to between~\SIrange{0}{10}{\volt}.  The
final signal is subsequently fed into a voltage-to-frequency
converter, and the output pulses sent to a set of counters for
acquisition and post-processing.

\begin{figure}
  \includegraphics[width=\columnwidth]{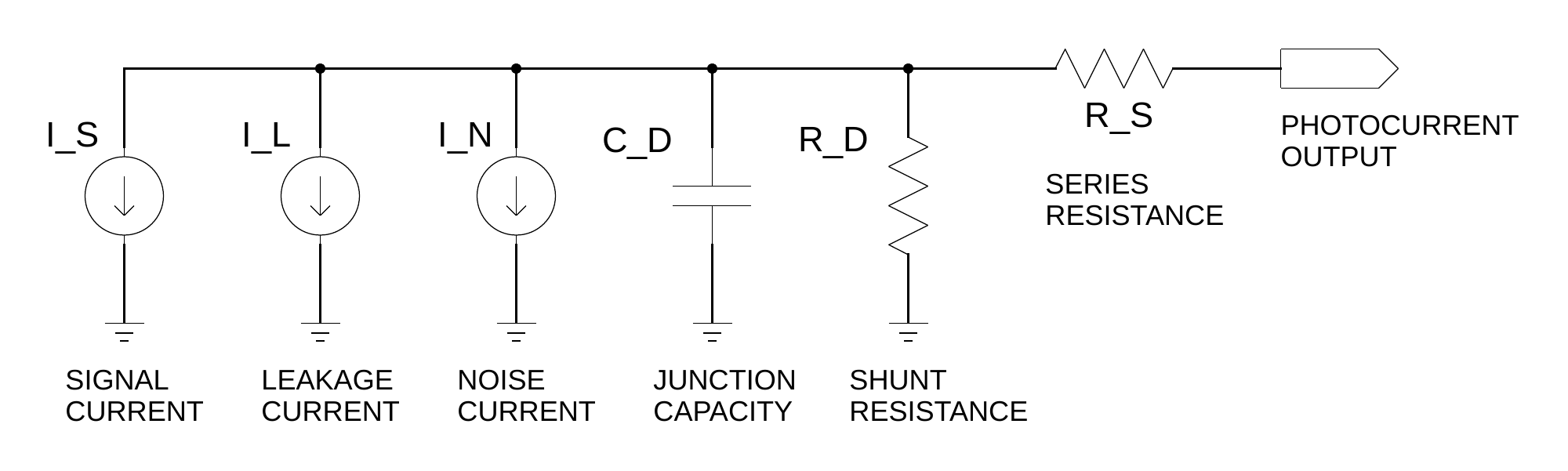}
  \caption{A photodiode can be modelled from a set of discrete
    components.}
  \label{fig:photodiode_model}
\end{figure}

When simulating this circuit, the photodiode is broken down into an
equivalent circuit of several discrete
components~\citep{graeme1996photodiode}, shown in
Figure~\ref{fig:photodiode_model}.  The model is simplified by
ignoring both the leakage current and noise current, since the
photodiode is operated in the low-noise photovoltaic mode of
amplification, and in a high-gain environment the diode intrinsic
noise becomes insignificant in comparison to that from the rest of the
system.  The series resistance is not stated by manufacturer, and so
this has been assumed to be low at~\SI{10}{\ohm}.  The photodiode has
a maximum junction capacitance of~\SI{2500}{\pico\farad}, and the
shunt resistance is typically~\SI{15}{\mega\ohm}.  The final
photodiode model is that of a signal current source, a parallel
capacitance, a parallel shunt resistance, and a series resistance.  We
can use this model along with SPICE~\citep{Nagel:M382,Nagel:M520}
simulation to analyse the performance of the BiSON detectors,
determining noise levels, frequency response, and finally slew rate.


\subsection{Noise Performance}

\begin{figure}
  \includegraphics[width=\columnwidth]{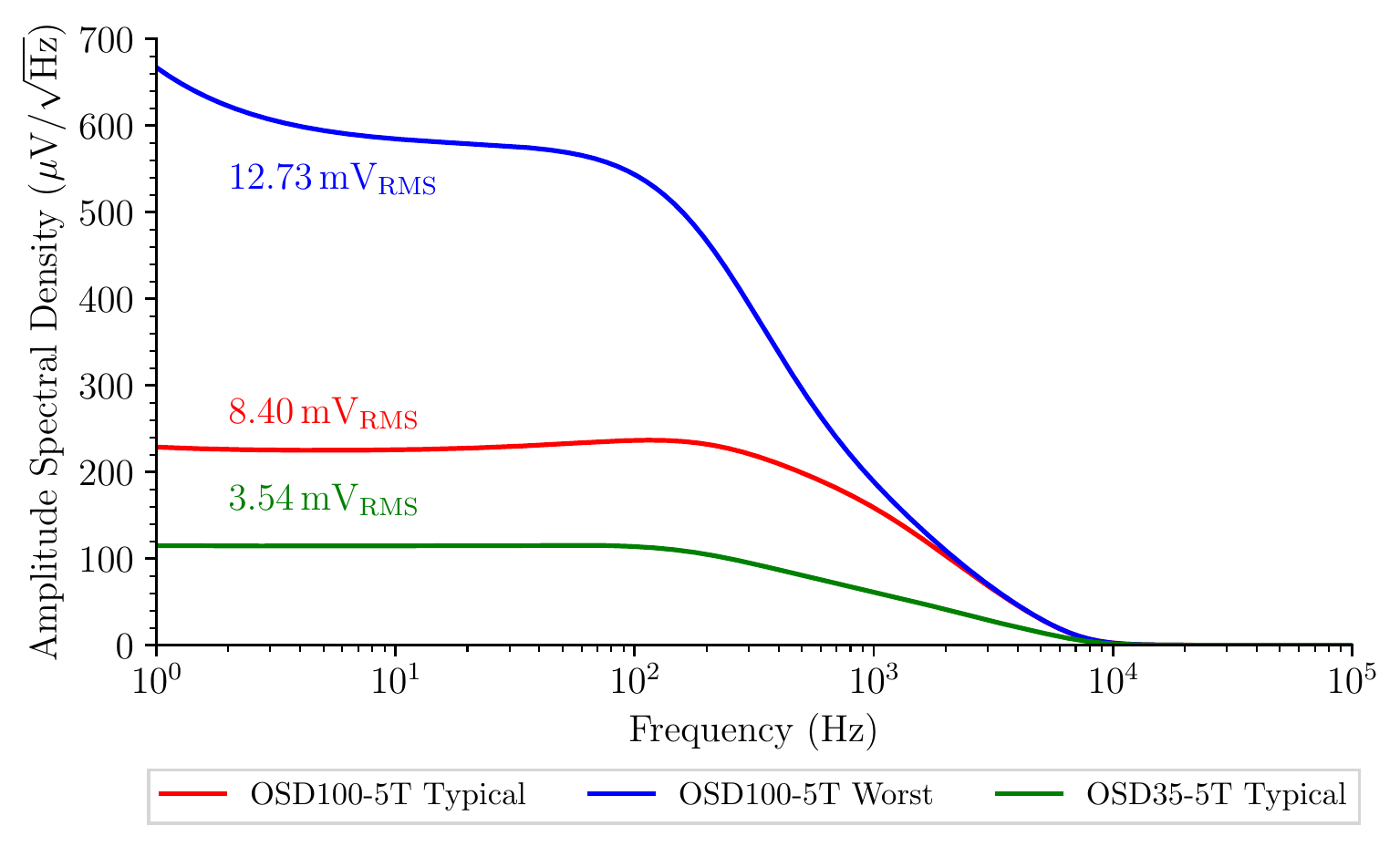}
  \caption{Scattering detector noise analysis.  Typical and
    worst-case performance is shown for the OSD100-5T photodiode based
    on the datasheet specification, and also for comparison the
    typical performance of a diode with a sensor of half the
    diameter.}
  \label{fig:jdet_noise}
\end{figure}

In a high-gain environment the noise level is generally dominated by
Johnson-Nyquist thermal noise from the gain resistance.
%
%
Over a~\SI{5}{\kilo\hertz} bandwidth at room temperature for
a~\SI{100}{\mega\ohm} gain resistor this
produces~\SI{0.09}{\milli\vrms} of thermal noise.  The detector has a
combined gain of~\num{62.4} after the initial current-to-voltage
conversion, and so this means we can expect to obtain
around~\SI{5.6}{\milli\vrms} of noise just from Johnson noise alone.


Another important source of noise in the circuit is produced by the
operational amplifiers.  The input bias current can become quite
significant depending on the overall gain of the system.  The input
bias current is also temperature dependent, and so a low temperature
coefficient is crucial for long term stability.
Figure~\ref{fig:jdet_noise} shows the results of the noise analysis,
using both typical and worst-case photodiode parameters provided by
the manufacturer datasheet, and also for comparison a smaller
photodiode.  The noise level is Johnson-Nyquist limited,
typically~\SI{8.4}{\milli\vrms}.  Assuming a typical detector voltage
output of~\SI{1}{\volt}, passing the noise through
equation~\ref{eq:ratio}, and applying a velocity sensitivity
of~\SI{3000}{\meter\per\second}, this results in an electronic
velocity noise level of the order of~\SI{2}{\centi\meter\per\second}
RMS.  The line-of-sight velocity component due to oscillations of the
Sun is of the order of~\SI{1}{\meter\per\second} and this is
superimposed on the much larger, up
to~\SI{1.5}{\kilo\meter\per\second}, components due to Earth's
rotation and the solar gravitational
redshift~\citep{1995A&AS..113..379E}.


\subsection{Frequency Response}

\begin{figure}
  \includegraphics[width=\columnwidth]{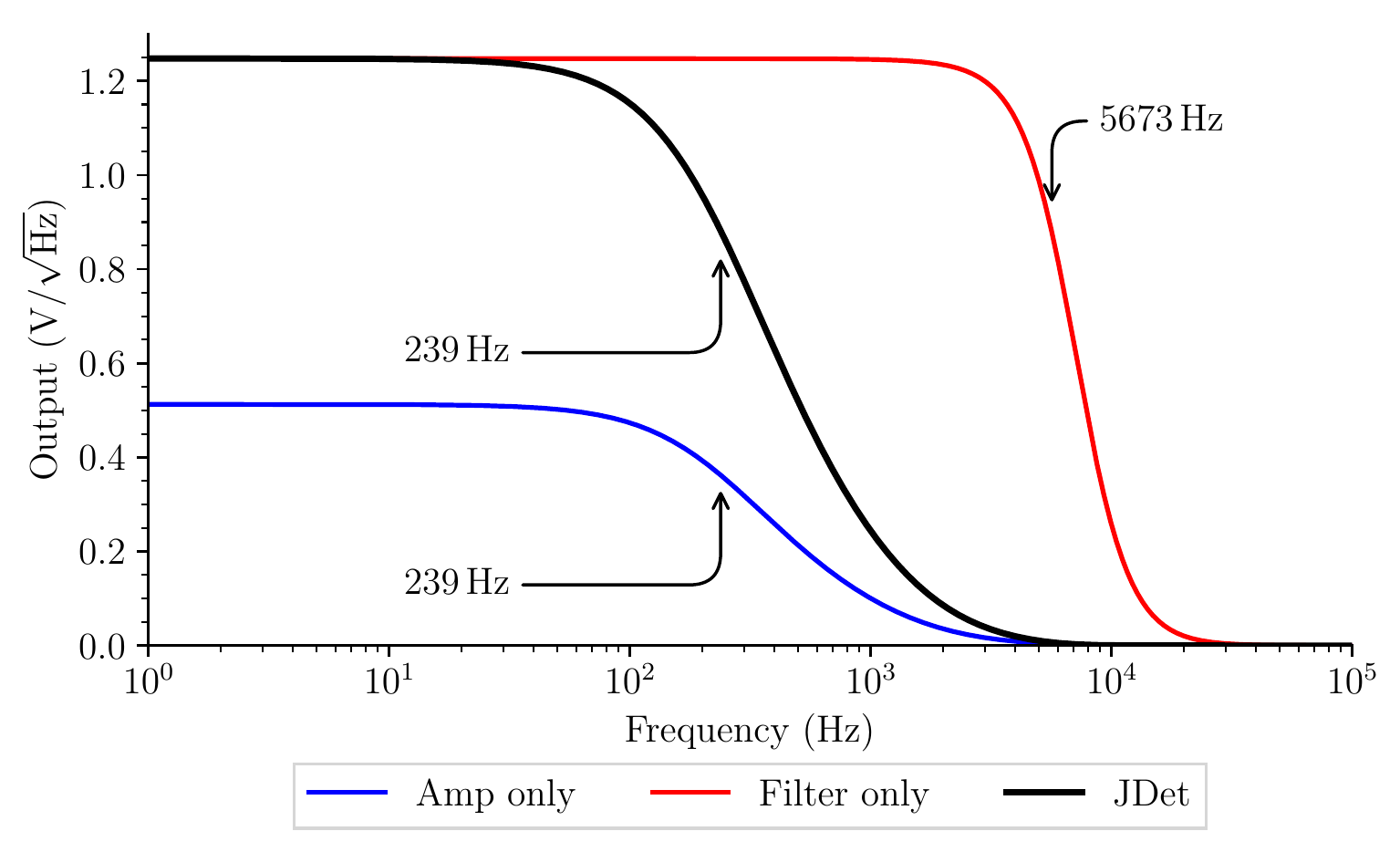}
  \caption{Scattering detector frequency response analysis.  The
    response of the initial twin-amplifier stage, and the low-pass
    filter stage, are shown separately.  The heavier weight black line
    shows the response of the whole detector.  The input signal
    is~\SI{200}{\pico\ampere} of alternating photocurrent.  The arrows
    indicate the~\SI{-3}{\decibel} point for each response curve.}
  \label{fig:jdet_response}
\end{figure}

The SPICE simulation of frequency response is shown in
Figure~\ref{fig:jdet_response}, including separately for both the
amplifier-stage, and the filter-stage.  Due to the very high gain of
the initial transimpedance amplifier the bandwidth is
just~\SI{239}{\hertz}.  The bandwidth of the Bessel low-pass filter is
actually higher at~\SI{5.6}{\kilo\hertz} than the bandwidth of the
first stage amplifier, and so the filter is doing nothing except
providing additional gain.  It is possible that the filter was
designed before the required gain and bandwidth were known.
%
%
Since we know the total noise is dominated by the bandwidth-limited
first stage transimpedance amplifier, the detector design could be
simplified by removing the filter completely and increasing the gain
of the second stage amplifier, eliminating more than half of the
components.


\subsection{Slew Rate and Transient Response}

\begin{figure}
  \includegraphics[width=\columnwidth]{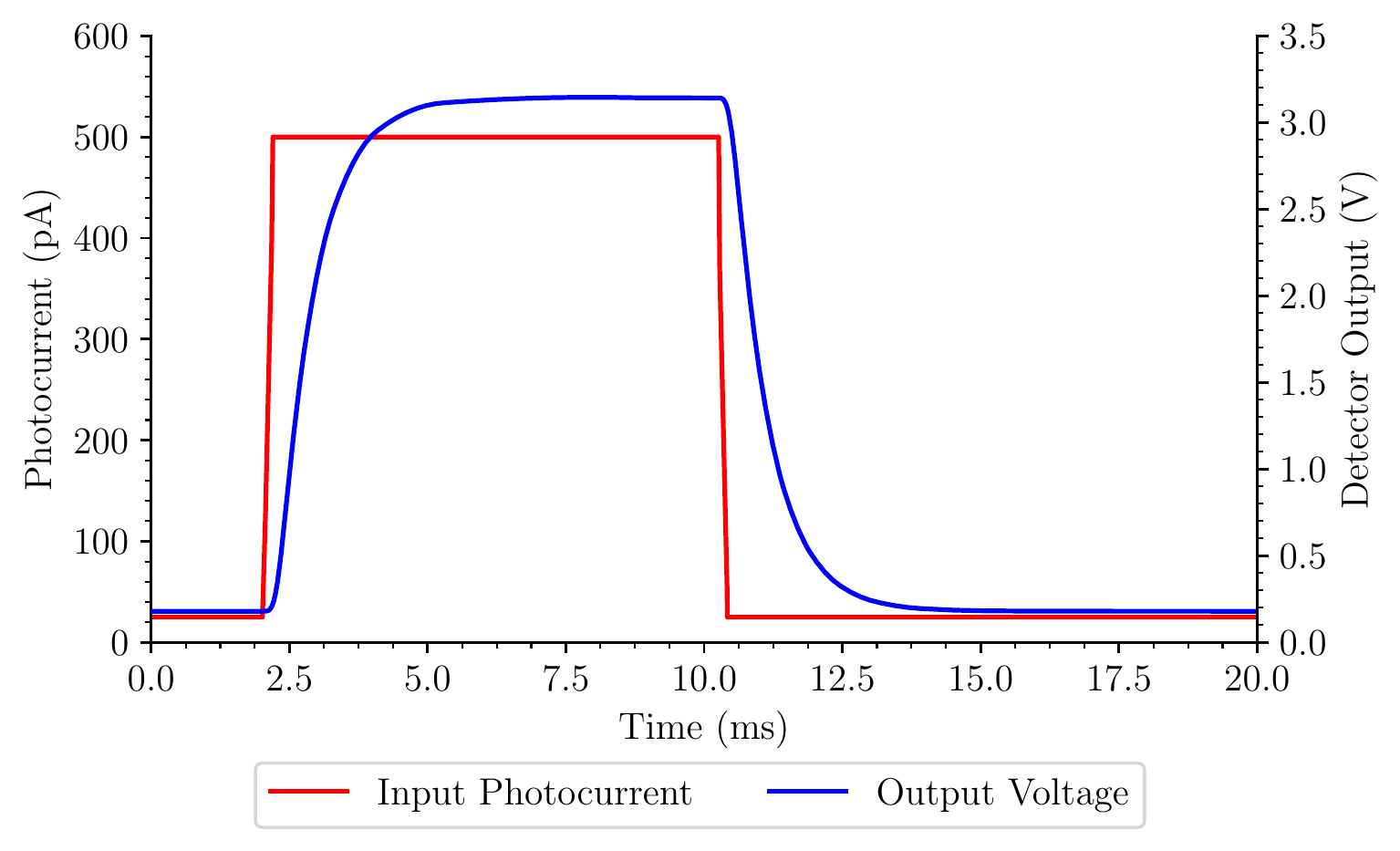}
  \caption{Scattering detector transient response analysis.  The
    transient response of the system is shown when stepping
    between~\SI{25}{\pico\ampere} of photocurrent
    and~\SI{500}{\pico\ampere} of photocurrent.}
  \label{fig:jdet_transient}
\end{figure}

A transient can be considered an ideal square wave.
%
%
The response of a circuit at high frequencies affects its processing
of short-time events.  \citet{johnson1993high} describe a rule of
thumb for the maximum practical frequency component as,
\begin{equation}\label{eq:fknee}
  f_\mathrm{knee} = \frac{0.5}{t_\mathrm{r}} ~,
\end{equation}
where $t_\mathrm{r}$ is the desired rise time of
a~\SIrange{10}{90}{\percent} transition, and $f_\mathrm{knee}$ the
inflection point or ``knee'' in the frequency spectrum above which
frequency components are insignificant in the determining the shape of
the signal.  If we require a \SIrange{10}{90}{\percent} rise time
of~\SI{0.5}{\milli\second}, then this suggests a bandwidth requirement
of~\SI{1}{\kilo\hertz}.  We know from the frequency response analysis
that the bandwidth is~\SI{239}{\hertz}, and so we can expect the rise
time to be longer than~\SI{2}{\milli\second} for
a~\SIrange{10}{90}{\percent} transition.
Figure~\ref{fig:jdet_transient} shows the SPICE transient response
analysis.  The simulated photocurrent is stepped
from~\SIrange{25}{500}{\pico\ampere}, resulting in a rise time
of~\SIrange{2.5}{4}{\milli\second} depending on the acceptable percent
of completion.


In the next section we look at two methods of electronically
controlling light polarisation state, and the slew rates that can be
achieved.


%
%
%
%


\section{Electro-optic Light Modulation}
\label{s:polarisation}


\subsection{Pockels-effect Cells}

Most BiSON instrumentation uses Pockels-effect cells constructed using
potassium dideuterium phosphate $\mathrm{KD_2PO_4}$ (KD*P), which
belongs to the tetragonal crystal
system~\citep{phillips2011introduction}.  This crystal exhibits
birefringence and is normally optically uniaxial, however when placed
in an electric field it becomes biaxial.  The KD*P crystal is z-cut,
which means that the two faces of the crystal are normal to the
crystallographic z-axis.  The on-axis rays through the
spectrophotometer are parallel to the crystallographic z-axis, and the
beam propagates in the positive z-direction.  The cell acts as a
linear retarder to a beam passing through the Pockels cell parallel to
the optical axis.  An electric field can be applied via transparent
electrodes applied to the input and output faces of the crystal.  The
retardance varies linearly with the voltage applied across the
electrodes, and at approximately~\SI{2300}{\volt} the retardance
reaches~\SI{45}{\degree} and the Pockels cell behaves like a
quarter-wave plate.  When combined with a linear polariser this causes
one of the polarisation components to be retarded by~\SI{90}{\degree}
in phase (a quarter of a wavelength) in relation to the other
component, producing circularly polarised light.  Circular
polarisation is considered to be left- or right-handed depending on
direction of rotation of the electric field vector of the wave.  If
the electric field across the cell is reversed then the retardance
becomes~\SI{-45}{\degree} and the opposite handedness of circular
polarisation is produced.


\begin{figure}
   \includegraphics[width=\columnwidth]{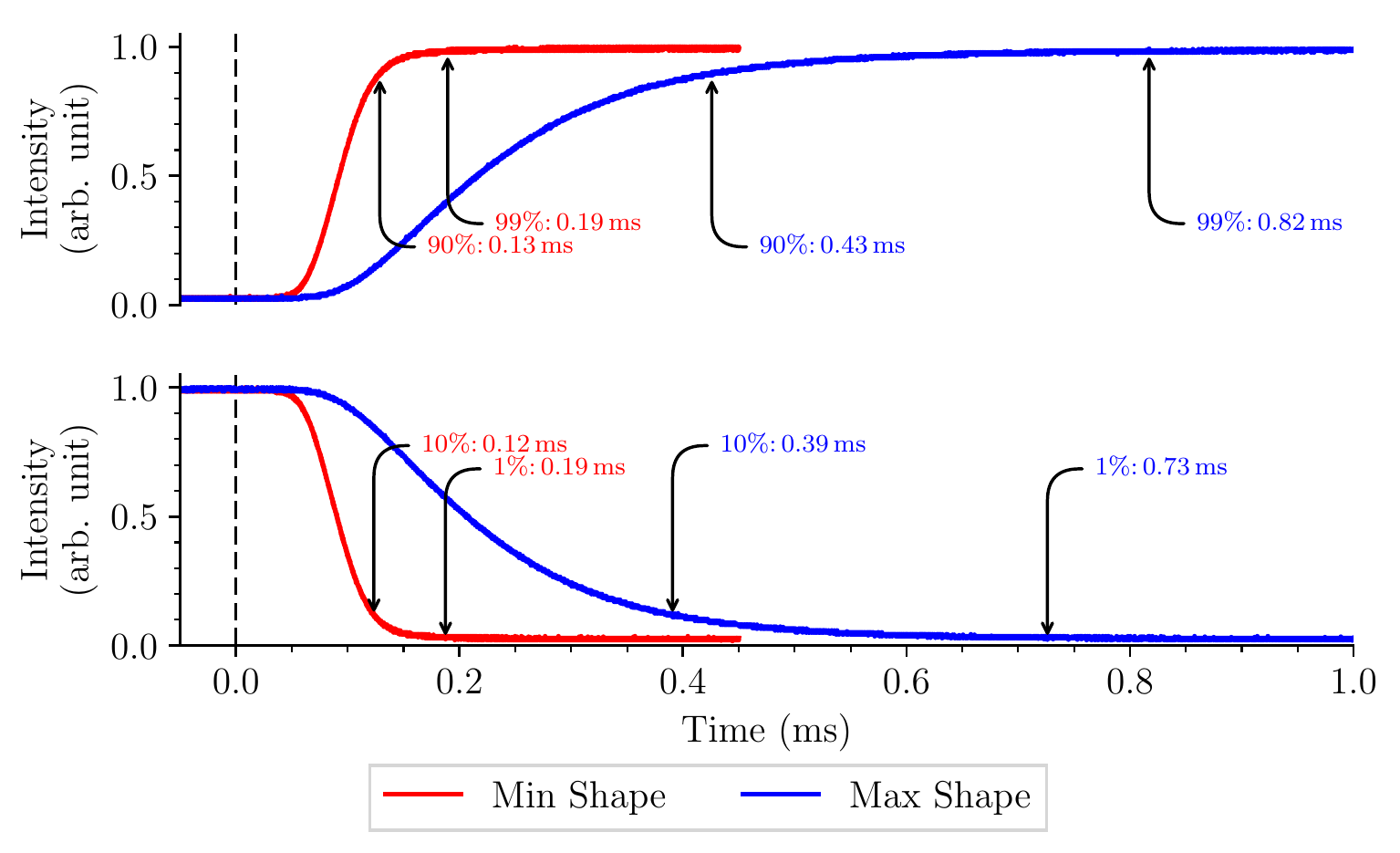}
   \caption{Pockels-effect cell switching rate.  The driver has a
     ``shape'' control that smooths the transition and helps to
     relieve mechanical stress on the crystal at the expense of a
     longer transition period.  The arrows indicate the time at which
     the stated percentage of the final value is reached.  The dashed
     vertical lines indicate the switching command trigger point.}
   \label{fig:pockels_rate}
\end{figure}

The switching rate of a Pockels cell was tested by monitoring the
change in light intensity between crossed polarisers.  In industry it
is typical to quote the slew-rate of a device as the time taken to
transition from~\SIrange{10}{90}{\percent} of the final value.  Here
we quote time taken from the instant the change is commanded to when
the output reaches~\SI{90}{\percent} and~\SI{99}{\percent} of the
final light intensity.  The switching rate for transitions in both
polarisation directions is shown in Figure~\ref{fig:pockels_rate}.
Power is supplied via a bespoke high-voltage supply.  The driver has a
``shape'' control that smooths the applied square-wave voltage
transitions and so helps to relieve mechanical stress on the crystal,
at the expense of a slightly longer transition period.  At minimum
smoothing the switching time to~\SI{99}{\percent} of the final
intensity is approximately~\SI{0.2}{\milli\second}.  With maximum
smoothing the switching time is approximately~\SI{0.8}{\milli\second},
and this is broadly symmetric in both directions.  The instrumentation
is typically operated with maximum smoothing to extend the life of the
crystal, and so the switching time is considered to be
nominally~\SI{1}{\milli\second}.

Both the field-widened Pockels cell and the required high-voltage
computer-controlled driver are bespoke devices that are difficult to
manufacture.  Whilst commercial Pockels cells are available, their
intended use is with highly collimated beams and so they have small
apertures and very limited field of view.  We will now look at a
commercial alternative that is a potential off-the-shelf direct
replacement.


\subsection{Liquid Crystal Retarder}

\begin{figure}
   \includegraphics[width=\columnwidth]{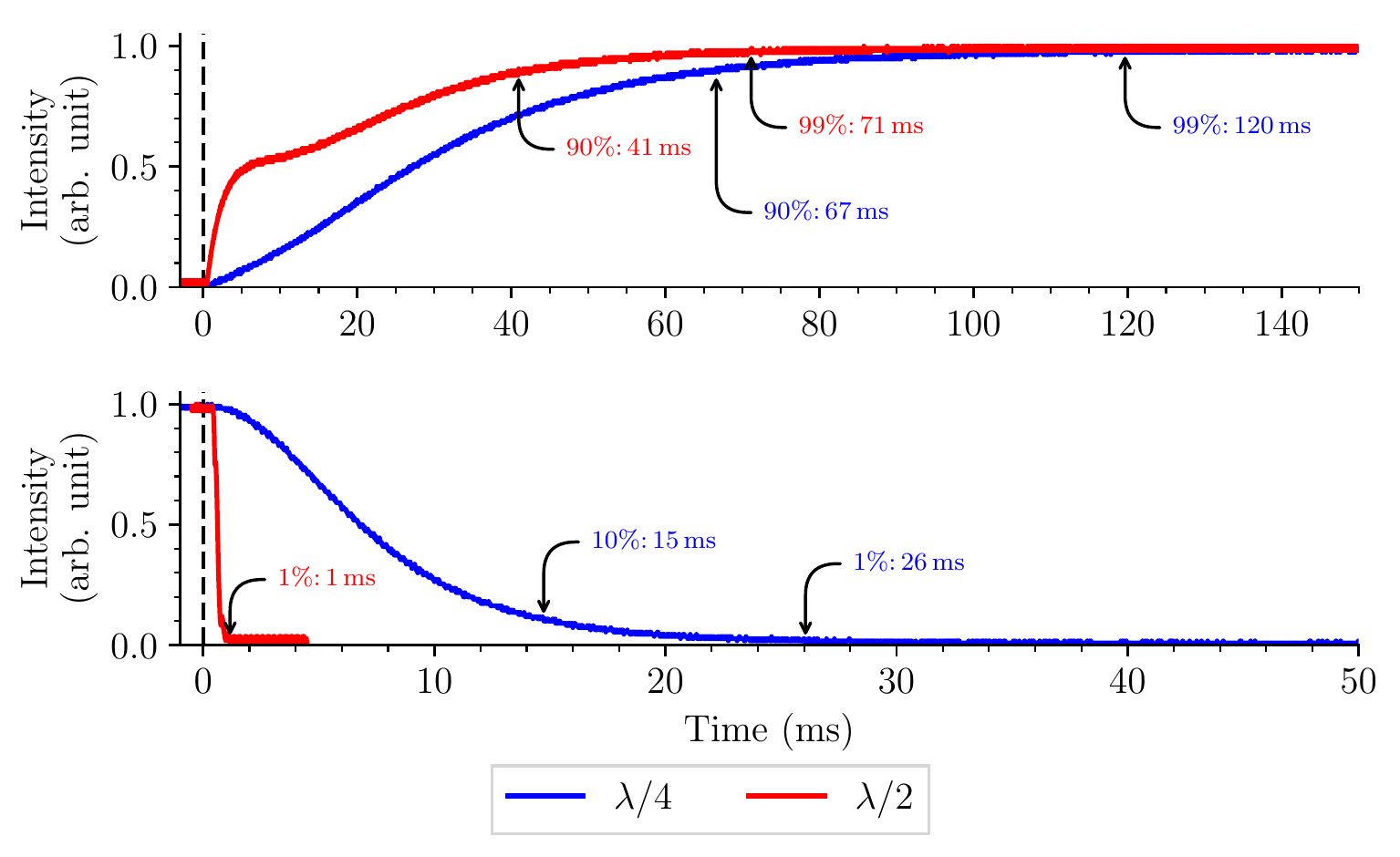}
   \caption{Liquid-crystal switching rate at room temperature.  Top:
     high-to-low voltage transitions.  Bottom: low-to-high voltage
     transitions.  The arrows indicate the time at which the stated
     percentage of the final value is reached.  The dashed vertical
     lines indicate the switching command trigger point.}
   \label{fig:lcd_rate}
\end{figure}

Many nematic crystal organic molecules are uniaxial with one long axis
and two others of equal length, similar to a cylinder.  In the nematic
phase the cylinders have no preferential order, but they self-align
with their long axes approximately parallel and this creates an
optical anisotropy.  When an electric field is applied the molecules
tilt and align to the field, and so changes in the field strength
cause changes in the effective retardance.  The driving signal is
typically between~\SIrange{0}{\pm25}{\volt}, considerably lower, more
convenient, and safer than the~\SI{\pm2300}{\volt} required by a
Pockels-effect cell.

However, LCD retarders are typically slower to switch between states
than a Pockels-effect cell.  Switching rate is affected by both the
size of the difference between applied voltages, and the crystal
temperature.  When switching from a low voltage to a high voltage, the
liquid crystal becomes under tension and the change occurs rapidly.
In the reverse direction the rate is limited by how long it takes the
cell to ``relax'' back to its natural state and this is determined by
the liquid viscosity, and so by the temperature.  The low-to-high
voltage transition is much faster than the high-to-low voltage
transition.  Higher temperatures and larger voltage differences allow
for faster switching.  The overall change in polarisation varies
slightly with temperature, and so it is important to calibrate the
control voltages at the intended operating temperature, and then
ensure stability at that temperature.

Figure~\ref{fig:lcd_rate} shows the time for both $\pm\lambda/4$ and a
$\pm\lambda/2$ transition at room temperature with times
to~\SI{90}{\percent} and~\SI{99}{\percent} indicated.  When
considering $\pm\lambda/4$ switching, the fast direction (low-to-high
voltage) takes between~\SIrange{15}{26}{\milli\second}. Switching in
the slow direction (high-to-low voltage) takes
between~\SIrange{67}{120}{\milli\second}.  The transition
between~$3\lambda/4$ and~$\lambda/4$ requires control voltages
of~\SI{1.78}{\volt} and~\SI{4.03}{\volt} respectively.

The switching time can be reduced by adding a fixed quarter-wave plate
and instead operating the LCD between zero- and half-wave retardance,
which requires control voltages of~\SI{25.0}{\volt} and~\SI{2.32}{\volt}
respectively.  This configuration reduces the fast transition to
just~\SI{1}{\milli\second} and the slow transition to
between~\SIrange{41}{71}{\milli\second}.  However, this remains much
slower than a Pockels-effect cell.

\begin{figure}
   \includegraphics[width=\columnwidth]{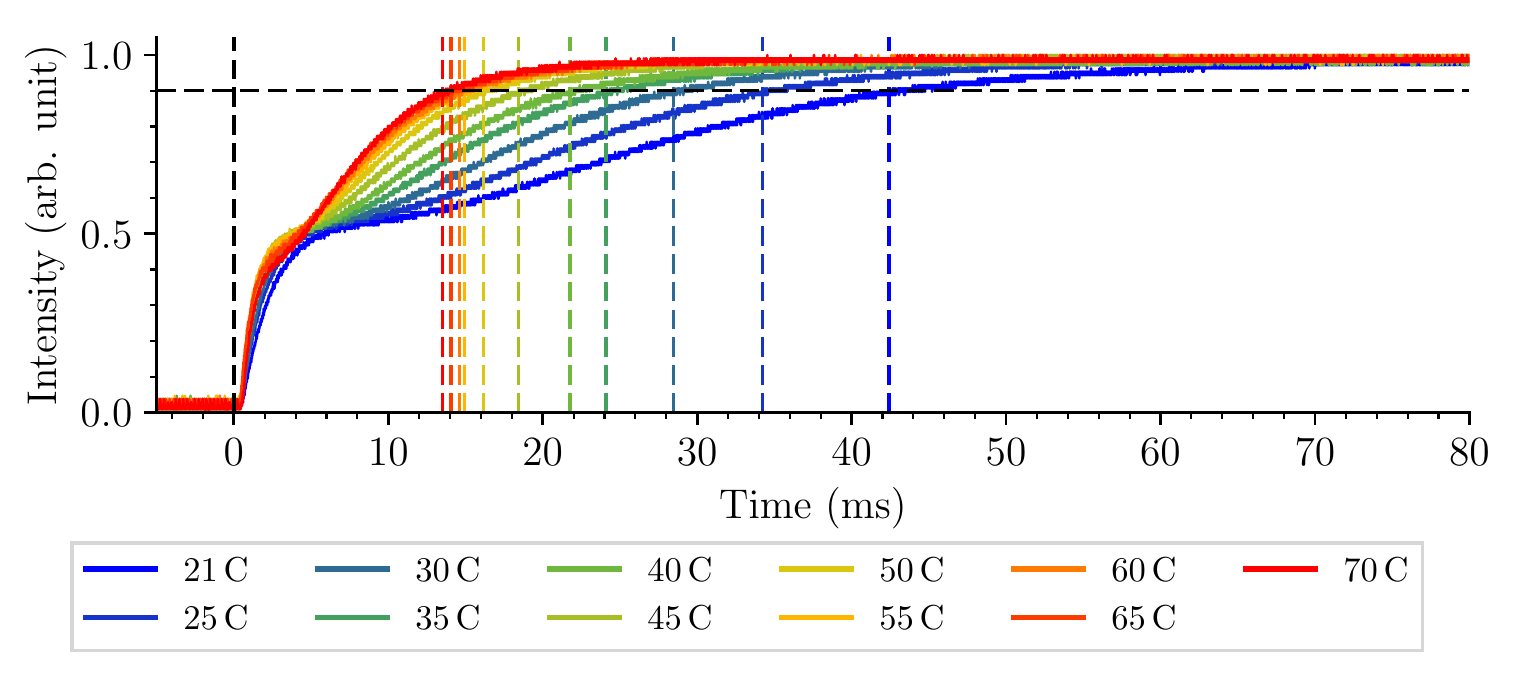}
   \caption{Temperature dependence of the LCD switching rate.  The
     black dashed vertical line indicates the switching command
     trigger point.  The dashed horizontal line indicates
     the~\SI{90}{\percent} complete point.}
   \label{fig:switching_temperature}
\end{figure}

The slower transition rate can be improved by heating the liquid
crystal.  Figure~\ref{fig:switching_temperature} shows the high-to-low
voltage switching time from zero-wave to half-wave for a range of
temperatures.  Table~\ref{table:lcd_temperatures} shows the switching
time for both the~\SI{90}{\percent} and~\SI{99}{\percent} transition
completion points.  At~\SI{50}{\celsius} the switching time is
between~\SIrange{16}{27}{\milli\second}.  There is little improvement
above~\SI{50}{\celsius} and so this is the ideal operating temperature
with respect to switching rate and maximising the life expectancy of
the crystal.

A transition time of~\SI{27}{\milli\second} is longer than that
achieved with a Pockel's effect cell.  We discuss the potential impact
of moving to LCD-based polarisation control in the next section.

\begin{table}
    \centering
    \caption[LCD switching rates at a range of temperatures]
    {LCD switching rates at a range of temperatures.}
    \label{table:lcd_temperatures}

    \begin{tabular}{r r r r r r r r r r r r}

    \toprule

    & \multicolumn{11}{c}{Switching rate (\si{\milli\second}) at Temperature (\si{\celsius})}\\
    \cmidrule{2-11}
    Temp & 21 & 25 & 30 & 35 & 40 & 45 & 50 & 55 & 60 & 65 & 70\\

    \midrule

    \SI{90}{\percent} & 42 & 34 & 28 & 24 & 22 & 18 & 16 & 15 & 15 & 14 & 14\\
    \SI{99}{\percent} & 74 & 57 & 48 & 42 & 43 & 31 & 27 & 26 & 27 & 27 & 25\\

    \bottomrule

    \end{tabular}

\end{table}


%
%
%
%


\section{Discussion}
\label{s:discussion}

The noise characteristics of resonant scattering spectrophotometers,
such as those deployed by BiSON, have been investigated by several
authors in terms of time-dependent effects on the solar oscillation
modes extracted from the observations --
see,~e.g.,~\citet{1976A&A....50..221G, 1978MNRAS.185....1B,
  1989A&A...222..361A, 1989ApJ...345.1088H, 1991SoPh..133...43H,
  2005MNRAS.359..607C}.  Here, we have considered electronic and
atmospheric effects on the whole frequency envelope.

It is known that atmospheric scintillation is the dominant noise
source in BiSON raw data~\citep{halephd}.  The BiSON
spectrophotometers mitigate noise from atmospheric scintillation by
rapidly switching the measurement wavelength between the red and the
blue wings of the solar D1~transition of potassium, with polarisation
switching rates targeting~\SI{100}{\hertz}.  The data acquisition
system applies a~\SI{5}{\milli\second} exposure time for each
polarisation state, and a~\SI{0.5}{\milli\second} stabilisation delay
between transitions, producing a switching rate of~\SI{90.9}{\hertz}.
This is repeated for \num{340}~cycles giving a total acquisition
period of~\SI{3740}{\milli\second}, of which ~\SI{3400}{\milli\second}
is exposure time.  The values are then read out and the cycle repeated
with a \SI{4}{\second}~cadence.  There is \SI{15}{\percent}~dead-time
due to stabilisation delays and data readout within each
\SI{4}{\second}~integration period.  We found here that the scattering
detector amplifier rise time is longer than expected, and requires up
to~\SI{4}{\milli\second} of stabilisation time resulting in some
polarisation state mixing between data channels which is reducing the
instrumental sensitivity.

When using NLC-based polarisation switching, the much longer rise time
of up to~\SI{27}{\milli\second} requires a slower switching rate to
avoid unacceptably high dead-time.  If we accept
a~\SI{20}{\milli\second} stabilisation period, then a switching rate
of~\SI{5}{\hertz} can be achieved with~\SI{20}{\percent}~dead-time.
This can be raised to~\SI{10}{\hertz} with~\SI{25}{\milli\second}
stabilisation and~\SI{50}{\percent}~dead-time.  Measurement of
scintillation noise at the BiSON sites has shown that
the~\SI{-10}{\decibel} point typically occurs
above~\SI{20}{\hertz}~\citep{2020PASP..132c4501H}, and so it would be
acceptable to reduce the switching rate a little if necessary.
Whilst~\SI{<10}{\milli\second} is ideal, up to~\SI{50}{\milli\second}
is acceptable.  However, the NLC performance remains slower than ideal
switching rates.  An FLC would be able to achieve performance
equivalent to a Pockels-effect cell, but at the requirement of a much
larger budget than NLC.

A prototype fibre-fed spectrophotometer based largely on off-the-shelf
components, and using NLC-based polarisation switching, has been
tested at the BiSON site in the
Canary~Islands~\citep{10.1093/rasti/rzac007}.  Whilst noise levels
are, as expected, slightly higher than the best BiSON sites, the
results are in-line with average network performance.  The
off-the-shelf nematic LCD retarder is a viable replacement for a
bespoke Pockels-effect cell.  The simplifications arising from the use
of off-the-shelf components allows easier and lower cost
instrumentation deployment, as utilised by BiSON:NG, the next
generation observing platform for
BiSON~\citep{halephd,10.1117/12.2561282}. Increasing the number of
contemporaneous observations from many sites allows noise to be beaten
down by combining many incoherent measurements of the observed noise
level~\citep{doi:10.1093/mnras/stx2177}, and this produces an overall
improvement in noise level for the network as a whole.


%
%
%
%


\section*{Acknowledgements}

We would like to thank all those who have been associated with BiSON
over the years.  We particularly acknowledge the invaluable technical
assistance at all our remote network sites. {BiSON} is funded by the
Science and Technology Facilities Council ({STFC}) grant ST/V000500/1.


%
%
%
%


\section*{Data Availability}

All data are freely available from the {BiSON} Open Data
Portal~--~\url{http://bison.ph.bham.ac.uk/opendata}.  Data products
are in the form of calibrated velocity residuals, concatenated into a
single time series from all {BiSON} sites.  Individual days of raw or
calibrated data, and also bespoke products produced from requested
time periods and sites, are available by contacting the authors.
Oscillation mode frequencies and amplitudes are available
from~\cite{2009MNRAS.396L.100B} and~\cite{2014MNRAS.439.2025D}.



\bibliographystyle{rasti}
\bibliography{references}




\bsp	
\label{lastpage}
\end{document}